\documentclass[intlimits,twoside,a4paper]{article}
\usepackage{amsmath,amssymb}
\usepackage{graphicx}
\usepackage{wrapfig}
\usepackage[T2A]{fontenc}
\usepackage[cp1251]{inputenc}

\usepackage{cmpj2}

\issue{2012}{15}{2}{23603}
\doinumber{10.5488/CMP.15.23603}

\title[Tethered chains in a binary mixture]%
{A density functional study of the structure of tethered chains in a binary mixture%
\thanks{Dedicated to Prof. Orest Pizio on the occasion of his 60th birthday.}}
\author[M. Bor\'owko, T. Staszewski]{M. Bor\'owko, T. Staszewski}
\address{Department for the Modeling of Physico-Chemical Processes, Maria Curie-Sk{\l}odowska University, \\ 20-031 Lublin, Poland}

\authorcopyright{M. Bor\'owko, T. Staszewski, 2012}
\date{Received January 20, 2012, in final form March 20, 2012}

\begin{document}
\maketitle
\begin{abstract}
A density functional study of the structure of a layer formed by chain molecules pinned to a solid surface is presented. The chains are modeled as freely joined spheres. Segments and all components interact via Lennard-Jones (12--6) potential.  The interactions of fluid molecules with the wall are described by the Lennard-Jones (9--3) potential.  We analyze how different parameters of the model affect the dependence of the brush height upon the mixture composition. We consider the effect of grafting density and the parameters characterizing the interactions of fluid molecules with the substrate and with the chains as well as interactions within the mixture. The changes in the brush height correlate with the adsorption of particular components.
\pacs 68.47.Mn, 61.25.H, 68.47.Pe, 82.35.Gh
\end{abstract}

\section{Introduction}

The solid surfaces modified with tethered chains have been extensively studied in recent years due to their importance for a wide range of technological processes, such as adhesion, colloidal stabilization, lubrication, drug delivery, nanotechnology and chromatography \cite{c1,c2,c3}. The theoretical studies included scaling theories~\cite{c4,c5}, self-consistent field methods~\cite{c6,c7,c8,c9,c9a,c9b,c9c,c9d}, single chain mean-field methods~\cite{c10,c11,c12}, density functional theories~\cite{c13,c14,c15,c16,c17,c18,c19,c20,c21} and computer simulations~\cite{c22,c23,c24,c25,c26,c27,c28,c29,c30,c31,c32,c33,c34,c35,c36,c37}. The structure of the polymer layer can play a fundamental role in these processes.

When free segments of tethered chains are not very strongly attracted by the solid surface, the structure of an end-grafted chain can be a ``mushroom'' or a ``brush'', depending on the conditions. In good solvents, the grafting density controls the structure of a polymer film. At low grafting densities, the chains are isolated and assume nearly unperturbed configuration. As the surface coverage increases, the chains overlap and the interchain repulsion causes the chains to extend away from the wall. The configuration switches in the brush-like structure.  In a real system the structure of the bonded phase is the result of a very complicated interplay between the entropic repulsion and attractive forces acting upon all the molecules. The problem of a change of the configuration of grafted chains with their length, the grafting density and the quality of solvent has been investigated~\cite{c1,c3,c4,c37}. The structure of a polymer brush immersed into a multicomponent solvent has been studied using an analytical self-consistent field theory~\cite{c9a,c9b,c9c,c9d}. The properties of the brush were determined based on the phase diagram of a polymer dissolved in a mixed solvent. A binary solvent with partially miscible components has been in the focus of these investigations. There has been found the effect of selective solvation of the immersed chains and big changes in their conformations. The studies have shown that the brush height depends on the solvent content in a complicated way. However, this approach has some restricting assumptions. One of them is the infinities of polymer chains. As a result, this theory does not describe the brush features on the scale less than the brush height~\cite{c9c}.

The surfaces modified with end-grafted chains are used as stationary phases in reversed-phase liquid chromatography. The most popular chemically bonded-phase consists of relatively short chains. Numerous studies have shown that the stationary phase takes an active part in chromatographic separations~\cite{c3}. This process is affected by the chemical nature of the bonded chains, their length, the number of chains attached to the supporting surface and the composition of hydro-organic mobile phases. The solvent-induced conformational changes of the bonded phases have been investigated using nuclear magnetic resonance~\cite{c38,c39}, fluorescence ~\cite{c40}, vibrational~\cite{c41,c42} and Raman~\cite{c43,c44} spectroscopic techniques. Martire and Boehm~\cite{c45} introduced the model of ``breathing'' surface in which the alkyl chains are swollen by solvent penetration in the presence of nonpolar solvents and collapse toward the wall in the presence of more polar solvents. The chromatographic stationary phases have been also studied using the molecular dynamics~\cite{c22,c23,c24,c34} and Monte Carlo method~\cite{c28,c29,c30,c31,c32}. The computer simulations do not support the hypothesis that the bonded chains collapse or elongate considerably with changes in the liquid phase composition. Under the assumed conditions only more subtle solution-induced changes in the structure of the bonded-phase were observed. However, these simulations have been performed for a few model systems only.  The effective screening of the space of the model parameters that alter the structure of the bonded phase would require long-standing computer simulations. Such model calculations can be efficiently carried out using the density theory.

In our previous paper~\cite{c46} the effect of strong adsorption of one-component fluid on the height of the bonded layer has been investigated. In this work we employ the density functional theory to describe the short grafted chains immersed by a mixture of two liquids. The main goal of this study is to investigate how the thickness of the bonded phase changes with the composition of the mixture. We consider the solvents with different affinities to the solid surface and the grafted chains.

\section{Theory}

We consider a binary mixture of spherical molecules in contact with a solid surface covered by grafted chains. We introduce the model described in our previous papers~\cite{c18,c19,c20}. The chains are tangentially joined $M$ spherical segments of the same diameter, $\sigma_\mathrm{c}$.  The chain connectivity is enforced by the following bonding potential
\begin{equation} \label{eq:1}
\exp \left[-\beta V_\mathrm{B}(\mathbf{ R})\right]=\prod_{i=1}^{M-1}\delta (|\mathbf{ r}_{i+1}-%
\mathbf{ r}_{i}|-\sigma_\mathrm{c})/4\pi (\sigma_\mathrm{c})^{2},
\end{equation}
where $\mathbf{ R}_k\equiv (\mathbf{r}_{1},\mathbf{r}_{2},\cdots ,\mathbf{r}%
_{M})$ is the vector specifying positions of segments, the symbol $\delta$ denotes
the Dirac function, $\sigma_\mathrm{c}$ is the polymer segment diameter
and $\beta^{-1}=k_\mathrm{B}T$. Each chain has a surface-binding segment located at its end (indexed as $i=1$) that is pinned to the wall by the following potential
\begin{equation} \label{eq:2}
\exp\left[-\beta v_{\mathrm{s}1}^{(\mathrm{c})}(z)\right]=C \delta (z-\sigma_\mathrm{c}/2),
\end{equation}
where $z$ is a distance from the surface, $C$ is a constant.  The remaining segments of the tethered chains ($i=2,3,\cdots,M$) are ``neutral'' with respect to the surface and they interact with the surface via the hard-wall potential
\begin{equation} \label{eq:3}
v_{\mathrm{s}i}=\left\{
\begin{array}{ll}
\infty & \qquad z < \sigma_\mathrm{c}/2, \\
0 & \qquad \text{otherwise}.
\end{array}
\right.
\end{equation}

The fluid molecules, however, interact with the surface via the Lennard-Jones (9--3) potential
\begin{equation} \label{eq:4}
v_{k\mathrm{s}}(z)=4\varepsilon_{k\mathrm{s}}\left[(z_0/z)^{9}-(z_0/z)^3\right],
\end{equation}
where $\varepsilon_{k\mathrm{s}}$  characterizes the strength of interaction between the
 $k$-th component and the adsorbent ($k=1,2$) and $z_0=\sigma_k/2$.

Interactions between segments and all fluid molecules are described by the Lennard-Jones (12--6) potential
\begin{equation} \label{eq:5}
u_{kl}(r)=\left\{
\begin{array}{ll}
4\varepsilon_{kl}\left[(\sigma_{kl}/r)^{12}-(\sigma_{kl}/r)^6\right]& \qquad r < r_\mathrm{cut}^{(kl)}, \\
0 & \qquad \text{otherwise},
\end{array}
\right.
\end{equation}
where $\sigma_{kl}=0.5(\sigma_k+\sigma_l)$ and the parameter  $\varepsilon_{kl}$ characterizes the interactions between species $k$ and $l$ $(k=\mathrm{c},1,2)$, $r_\mathrm{cut}^{(kl)}=5\sigma$.

We split these interactions into the repulsive and attractive parts according to the Weeks-Chandler-Anderson scheme~\cite{c47}
\begin{equation} \label{eq:6}
u^{(kl)}_\mathrm{att}(r)=\left\{
\begin{array}{ll}
-\varepsilon_{kl} &  r <  2^{1/6} \sigma_{kl}\,, \\
u_{kl}(r)& r \geqslant 2^{1/6} \sigma_{kl}\,.
\end{array}
\right.
\end{equation}

We define the total segment density of the grafted chains as follows:
\begin{equation}  \label{eq:7}
\rho _\mathrm{s}^{(\mathrm{c})}(\mathbf{ r})=\sum_{i=1}^{M}\rho _{\mathrm{s},i}^{(\mathrm{c})}(\mathbf{ r}%
)=\sum_{i=1}^{M}\int \rd\mathbf{R}\delta (\mathbf{r}-\mathbf{ r}%
_i)\rho^{(\mathrm{c})}(\mathbf{ R})\;,
\end{equation}
where $\rho_{\mathrm{s},i}^{(\mathrm{c})}(\mathbf{r})$ is the density of the individual i-th segment and $\rho^{(\mathrm{c})}(\mathbf{R})$ is the total density of grafted chains.

In the system the condition of constancy of the number of grafted molecules is fulfilled
\begin{equation} \label{eq:8}
\int_0^{(M+1/2)\sigma_\mathrm{c}} \rd{ z}\rho_{\mathrm{s},i}^{(\mathrm{c})}(z) =\rho_\mathrm{c}\,,
\end{equation}
where $\rho_\mathrm{c}$ is the grafting density defined as $\rho_\mathrm{c}=N_\mathrm{c}/A$, $N_\mathrm{c}$ is the number of chains, A denotes the surface area.
For such a system the thermodynamic potential is given by
\begin{equation} \label{eq:9}
{ Y} = F\left[\rho ^{(\mathrm{c})}(\mathbf{ R}),\rho_1(\mathbf{ r}_1),\rho _2(\mathbf{ r}_2)\right]
+ \int\rd\mathbf{ R}\rho^{(\mathrm{c})}(\mathbf{R})v^{(\mathrm{c})}(\mathbf{ R})
+ \sum_{k=1,2} \int\rd\mathbf{ r}_k\rho_k(\mathbf{ r}_k)[v_k(\mathbf{ r}_k)-\mu_k],
\end{equation}
where $\mu_k$ is the chemical potential of the k-th component.

We describe the system in the framework of the density functional theory proposed by Yu and Wu~\cite{c48,c49,c50}. In our previous papers we applied the theory to the study of adsorption on the bonded-phases~\cite{c18,c19,c20}. Since the details of computational method have been already described elsewhere we outline here only most important steps of the calculations.

As usual, we split the free energy functional into the sum, $F=F_\mathrm{id}+F_\mathrm{hs}+F_\mathrm{c}+F_\mathrm{att}$.
The ideal part of the free energy ($F_\mathrm{id}$) is known exactly~\cite{c51}. The hard sphere contribution ($F_\mathrm{hs}$) is calculated according to the fundamental measure theory of Rosenfeld~\cite{c52}. The excess free energy  due to the chain connectivity ($F_\mathrm{c}$) follows from the first order perturbation theory of Wertheim~\cite{c53}.  The attractive interactions between molecules of the liquid mixture and their interactions with the chain segments are described using a mean field approximation. One can find  appreciable formulas in reference~\cite{c21} (equations~(7), (10), (12), (14)).
Next, we minimize the thermodynamic potential (\ref{eq:9}) under the constraint (\ref{eq:8}). As a result, we obtain a set of Euler-Lagrange equations (equation~(17) in~\cite{c21}) and solve it numerically.
This procedure provides the density profiles of all species. In our model all densities depends only on the distance from the wall, i.e, $\rho^{(\mathrm{c})}_\mathrm{s}=\rho^{(\mathrm{c})}_\mathrm{s}(z)$ and $\rho_k=\rho_k(z)$.

We calculate the brush height from the first moment of the total segment density profile~\cite{c37}
\begin{equation} \label{eq:10}
h =2\frac{\int \rd z z\rho_\mathrm{s}^{(\mathrm{c})}(z)}{\int \rd z \rho_\mathrm{s}^{(\mathrm{c})}(z)}\,.
\end{equation}

Next, we can evaluate the excess adsorption isotherms of particular components
\begin{equation}  \label{eq:11}
\Gamma_k = \int \left [ \rho_k(z)-\rho_{k\mathrm{b}} \right ]\rd z
\end{equation}
and the relative excess adsorption isotherm defined as
\begin{equation}  \label{eq:12}
N^\mathrm{e}_k = \int \left[ x_k(z)-x_{k\mathrm{b}} \right]\rd z,
\end{equation}
where $x_k=\rho_k/\rho_\mathrm{F}$ is the local molfraction
 $k$-th component and $x_{k\mathrm{b}}$ is its value in the bulk phase whereas $\rho_\mathrm{F}=\rho_1+\rho_2$ is the total  density of the fluid.

The brush height is usually scaled by the radius of gyration of the chain molecules~\cite{c37}. Unfortunately, we cannot obtain the radius of gyration within the considered theory. Therefore, we used reduced units which are defined below.

We have assumed that the segments of the chains and all spherical molecules have the same diameters $\sigma_k=\sigma$, $(k=\mathrm{c},1,2)$ and take $\sigma$ as the unit of length. The energy parameter characterizing interactions between molecules of the component 2 has been treated as the unit of energy,  $\varepsilon_{22}=\varepsilon$. We have introduced the usual reduced unit as $\varepsilon_{kl}^*= \varepsilon_{kl}/ \varepsilon$ and $\varepsilon_{k\mathrm{s}}^*=\varepsilon_{k\mathrm{s}}/\varepsilon$,  ($k,l=\mathrm{c},1,2$). Similarly, the reduced densities are defined as $\rho^*_k(z)= \rho_k(z) \sigma^3$, $\rho^*_{k\mathrm{b}}(z)= \rho_{k\mathrm{b}}(z) \sigma^3$,
$\rho^{(\mathrm{c})*}_{\mathrm{s},i}(z)= \rho^{(\mathrm{c})}_{\mathrm{s},i}(z) \sigma^3$, $\rho_\mathrm{s}^{(\mathrm{c})*}=\rho_\mathrm{s}^{(\mathrm{c})}\sigma^3$ and $\rho_\mathrm{c}^*=\rho_\mathrm{c}\sigma^2$.  The reduced temperature is defined as $T^*=kT/\varepsilon$.

\section{Results and discussion}

In our model the parameter $\varepsilon_{2\mathrm{s}}^*$ characterizes the interactions of the k-th component with the solid surface. The affinity of the k-th component to the grafted chains is altered by adjusting the value of the parameter $\varepsilon _{k\mathrm{c}}^*$. The interactions in the bulk mixture are characterized by the parameters $\varepsilon _{11}^*$, $\varepsilon _{22}^*$ and $\varepsilon _{12}^*$. Depending on the values of these parameters the bulk mixture can undergo a demixing transition. However, all the fluids considered here exhibit a complete mixing in the bulk phase.

The system in question depends on numerous parameters. In order to reduce the number of parameters to a minimum, we assume that $\sigma_{kl}=\sigma$ and $\varepsilon _{22}^*= \varepsilon _\mathrm{cc}^*=\varepsilon_{2\mathrm{s}}^*=1$. All calculations have been carried out at $T^*=1$ and for the fluid bulk density $\rho_\mathrm{F}^*=0.66$. This density nearly coincides with the liquid density on the liquid coexistence curve at $T^*=1$.

The goal of the study is to estimate the effect of the selected parameters on the height of the brush. We have focused our attention on the role of the following factors: (i) the interactions of the both components with the bare surface (the surface effects), (ii) the interactions of the fluid molecules with segments of the grafted chains and (iii) the interactions in a bulk mixture (the solvent effects).

\begin{figure}[ht]
\centerline{\includegraphics[width=0.55\textwidth]{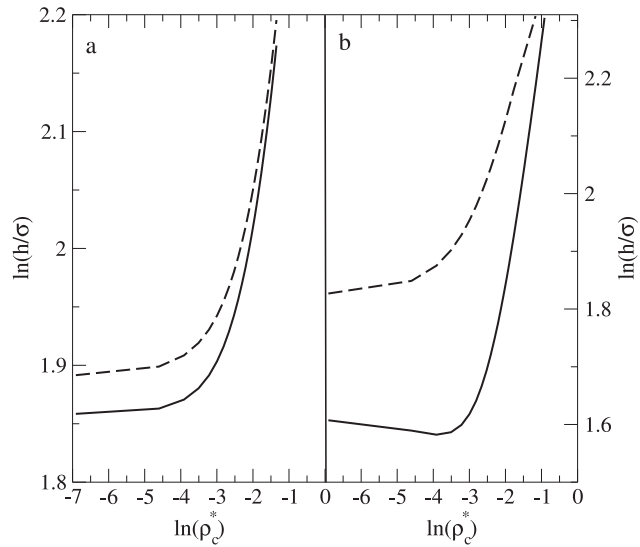}}
\caption{The dependence of the brush height on the grafting density of tethered 18-mers for two values of the molfraction: $x_{1\mathrm{b}}=0.2$ (solid lines) and $x_{1\mathrm{b}}=0.8$ (dashed lines). In all cases  $\varepsilon^*_{1\mathrm{s}}=15$, the parameter characterizing interactions with the chains $\varepsilon^*_{1\mathrm{c}}=1$ (part (a)) and $\varepsilon^*_{1\mathrm{c}}=1.1$ (part (b)).} \label{fig1}
\end{figure}

In figure~\ref{fig1} we demonstrate the effect of the selected parameters of the model on the relationship of the logarithm of the brush height upon the logarithm of the grafting density. The curves shown in part~(a) have been obtained for
a simple ``adsorption system'', for which interactions of molecules of the both components in the liquid and their interactions with the brush are identical, $\varepsilon^*_{kl}= \varepsilon^*_{k\mathrm{c}}=1, (k, l=\mathrm{c}, 1, 2)$ whereas the strengths of interactions with the wall are different and the energy parameters are equal to $\varepsilon^*_{1\mathrm{s}}= 15$ and  $\varepsilon^*_{1\mathrm{s}}=1$. In this system the driving forces for adsorption are the interactions of fluid molecules with the underlying surface. In part (b) we present the relationships  $\ln(h)~\text{vs}~ \ln(\rho_\mathrm{c}^*)$ for the model systems in which the first component has additionally a high affinity to the grafted chains, $\varepsilon^*_{1\mathrm{c}}=1.1$. The remaining parameters are the same as in part (a). In the both figures we show the log-log plots of the brush height as a function of the grafting density calculated for different mixture compositions, $x_{1\mathrm{b}}=0.2$ (solid lines) and $x_{1\mathrm{b}}=0.8$ (dashed lines).
All curves have the same shapes. When the number of tethered chains is small, i.e. at low surface coverages, the brush height is almost unaffected by the grafting density. Under such conditions the grafted chains do not practically effect one another and they assume unperturbed configurations. For mediate-covered surfaces a minimum on the curve is observed for certain parameters (see, the solid line in part (b)). With a further increase of the grafting density the brush height increases markedly. In this region the logarithm of the brush height changes with $\ln \rho_\mathrm{c}^*$  almost linearly.  A similar behavior was observed in computer simulations~\cite{c34}.

One sees in the figure~\ref{fig1} that the composition of the liquid mixture strongly affects the brush height at low surface coverages.  The brush height increases with an increase of the molfraction $x_{1\mathrm{b}}$. The same effect gives an increase of the parameter $\varepsilon_{\mathrm{c}1}$ (cf., dashed lines in parts (a) and (b)). Accumulation of the fluid molecules close to the wall causes an increase in the brush height because the liquid molecules compete for space inside the bonded-phases and the chains are pushed away from the surface.

In the previous paper~\cite{c46} we have studied the changes in the structure of tethered chains in contact with a one-component fluid. We have shown that in the region of  mediate grafting coverages, a minimum on the curve $\ln(h)~\text{vs}~\ln(\rho^*_\mathrm{c})$ can develop. For a strongly attractive surface this minimum disappeared. The minimum also vanishes when segment-segment attraction is lowered or when temperature is raised.  The similar effects are observed for the grafted chains immersed by the liquid mixture. In part (b) one sees the minimum on the curve corresponding to the low molfraction of the first component (solid line). However,
when the concentration of the first component in the bulk mixture increases, the minimum disappears. In this case molecules of the first component are strongly attracted by the substrate as well as by the grafted chains. Thus, we observe more molecules 1 in the mixture, higher accumulation of the fluid molecules in the surface layer which inhibits the chain coiling.

\begin{wrapfigure}{i}{0.5\textwidth}
\centerline{\includegraphics[width=0.48\textwidth]{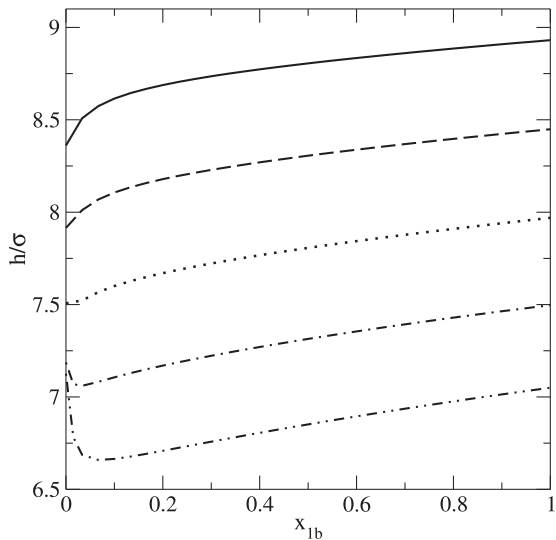}}
\caption{The dependence of the brush height on the molfraction of the first mixture component for different grafting densities $\rho_\mathrm{c}^*$: $0.05$ (double dotted-dashed line), $0.1$ (dot-dashed line), $0.15$ (dotted line), $0.2$ (dashed line), $0.25$ (solid line). The grafted chains are 18-mers. The calculations were carried out for $\varepsilon^*_{1\mathrm{s}}=15$ and $\varepsilon^*_{1\mathrm{c}}=1$.} \label{fig2}
\vspace{3mm}
\end{wrapfigure}

At high grafting densities the following relation is fulfilled
\begin{equation}  \label{eq:13}
\ln(h)\sim \gamma \ln(\rho_\mathrm{c}^*).
\end{equation}

We have found that exponent $\gamma$ depends on the mixture composition and on the parameters characterizing the system. For the ``adsorption'' system (part (a)) we have obtained $\gamma(x=0.2)=0.26$ and $\gamma(x=0.8)=0.24$. However, for a stronger attractive interaction of the first component with the chains (part (b)) the exponents are $\gamma(x=0.2)=0.48$, $\gamma(x=0.8)=0.21$

\begin{figure}[ht]
\centerline{\includegraphics[width=0.55\textwidth]{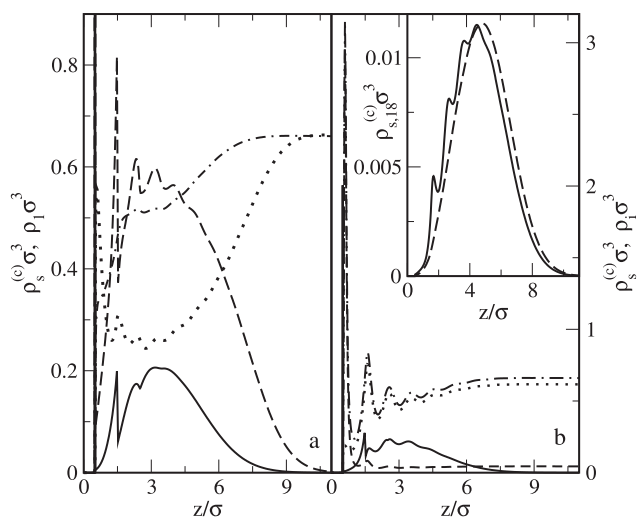}}
\caption{The segment density profiles and the density profile of the
mixture components. Part (a) is for the brush in contact with the pure
component 2 ($x_{1b}=0$) and two grafting densities $\rho_c^*$: $0.05$
(the segment density profile -- solid line, the fluid density profile -- dot-dashed line) and $0.2$  (the segment density profile -- dashed line,
the fluid density profile -- dotted line). Part (b) is for the mixture, in
which  $x_{1b}=0.0667$ and the grafting density $\rho_c^*=0.05$ .  The
density profiles are plotted for: the brush (solid line), the first
component (dashed line), the second component (dotted line) and for the
total density of the fluid (dot-dashed line). In the inset the density
profiles of end-segments are shown for $x_{1b}=0$ (dashed line) and
$x_{1b}=0.0667$ (solid line). The calculations were carried out for
18-mers and $\varepsilon^*_{1s}=15$ $\varepsilon^*_{1c}=1$.} \label{fig3}
\end{figure}

Figure~\ref{fig2} illustrates how the brush height changes with the mixture composition for selected grafting densities. The remaining system parameters are the same as in figure~\ref{fig1}. For sufficiently dense bonded-phases the brush height is a monotonically increasing function of the mole fraction $x_{1\mathrm{b}}$ in the whole concentration region. When  $x_{1\mathrm{b}}>0.15$ the brush height is almost a linear function of the mixture composition. An interesting feature is observed for low concentrations of the first component. For high grafting densities the brush height rapidly increases. The opposite effect is observed at low surface coverages. After adding a small amount of the first component to a pure component 2, the brush height decreases until a minimum value is attained. A further increase of the molfraction $x_{1\mathrm{b}}$ causes an increase of the brush height. The analysis of density profiles in a pure fluid 2 and in a mixture allows us to understand this effect (see figure~\ref{fig3}).

Figure~\ref{fig3}~(a) shows an effect of grafting density on the structure of the surface layer in a pure component 2 that weakly interacts with the solid surface ($\varepsilon^*_{2\mathrm{s}}=1$).  We start with a comparison of the segment density profiles of the grafted chains calculated at two grafting densities $\rho_\mathrm{c}^*=0.05$ and $\rho_\mathrm{c}^*=0.2$.  At very low density, $\rho_\mathrm{c}^*=0.05$, only two local peaks are observed on the segment density profile. However, for a denser bonded-phase there are four well-pronounced peaks, which correspond to successive layers of polymer segments. An extended region, where the segment density smoothly tends to zero, follows these peaks. For a higher grafting density the chains are more stretched, so the brush height is greater than that for a low grafting density. Also, density profiles of the fluid in the considered systems are different. In the both cases, depletion in the fluid density is observed in the surface layer. At low grafting density, the fluid density gradually decreases when the distance from the wall decreases. There is a local minimum close to the surface where the minimum fluid density is much lower than the bulk fluid density. A distinct behavior is observed for a higher coverage $\rho_\mathrm{c}^*=0.2$. There are three peaks at local density of the fluid. Their positions are correlated with the positions of the peaks at the segment density profiles of the grafted chains. The height of the local density peak adjacent to the wall is lower than the bulk density of the fluid.  The grafted chains attract molecules of the fluid. However, the fluid density in the middle part of the surface layer is considerably lower for the denser bonded-phase. On the contrary, close to the wall the fluid density for a low grafting density is lower than that for a dense brush. The grafted chains are always a space barrier for the fluid molecules. However, in the dense bonded-phase there are ``tunnels'' between the stretched chains,  and fluid molecules can flow through them to the wall. Due to the balance between attractive interactions and the entropic repulsion, the primary adsorption, i.e., adsorption near the surface is preferred.

Now, we analyze the structure of the surface layer for $\rho_\mathrm{c}^*=0.05$ after adding a small amount of the component 1 to the pure component 2. The density profiles presented in Figure 3b are calculated at the molfraction of the first component $x_{1\mathrm{b}}=0.0667$, which corresponds to the minimum brush height (see double dot-dashed line in figure~\ref{fig2}). The density profile of the second component is almost the same as that in the pure fluid 2.  However, the presence of the molecules 1 changes the total density profile of the fluid. The molecules 1 are strongly attracted by the solid surface. For this reason, close to the wall the density of the first component is markedly greater than in the bulk mixture. As a consequence, also the total density of the fluid is high in this region, whereas the fluid density far away from the surface only slightly differs  from the bulk value. Notice that at such a surface coverage the tethered chains are isolated  from one another but the molecules 1, which have been adsorbed near the wall, attract them. This causes the chain segments to cumulate somewhat closer the surface and we observe a decrease of the brush height. The results shown in the inset confirm this conclusion. We present here the density profile of the end-segment of the grafted chains for $x_{1\mathrm{b}}=0$ and $x_{1\mathrm{b}}=0.0667$. One sees that the end-segment density close to the surface is lower for $x_{1\mathrm{b}}=0$. This means that the chains are less coiled in the pure component 2 than in the considered mixture. Moreover, the end-segment density profile has a maximum at the distance from the surface at the points $z=4.80$  and $z=4.45$, respectively for $x_{1\mathrm{b}}=0$ and $x_{1\mathrm{b}}=0.0667$. The situation changes when  concentration of the first component increases. Then, the fluid density  considerably increases in the vicinity of the surface. The fluid molecules start to compete with the grafted chains for place in the surface layer. The entropic repulsion prevails the attraction between fluid molecules and chain segments. The chains are pushed away from the surface by fluid molecules and the brush height increases.

\begin{wrapfigure}{i}{0.5\textwidth}
\centerline{\includegraphics[width=0.48\textwidth]{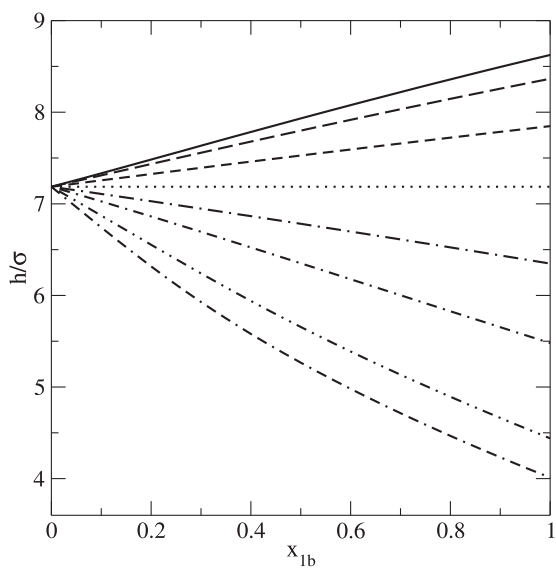}}
\caption{The dependence of the brush height on the molfraction of the first mixture component for different values of the parameter $\varepsilon^*_{1\mathrm{c}}$ characterizing the interactions of the molecules 1 with the brush. From the top: 1.13, 1.1, 1.05, 1.0, 0.95, 0.9, 0.8, 0.7. The grafted chains are 18-mers. The calculations are carried out for $\varepsilon^*_{1\mathrm{s}}=1.0$ and $\rho_\mathrm{c}^*=0.1$.} \label{fig4}
\end{wrapfigure}

Next, we discuss an impact of the solvent nature on the structure of the surface layer. In this order we have carried out the calculations assuming that interactions of the both components with the surface are the same and rather weak, namely $\varepsilon^*_{1\mathrm{s}}=\varepsilon^*_{2\mathrm{s}}=1$. First, we have studied the effects of the brush-fluid interactions. We have varied the parameter $\varepsilon_{1\mathrm{c}}^*$ and kept the remaining parameters fixed  $\varepsilon^*_{2\mathrm{c}}=\varepsilon^*_{kl}=1$ for  $k,l=1,2$. These results are presented in figure~\ref{fig4}. One sees here that when the component 1 is a better solvent for the chains than the component 2, $\varepsilon^*_{1\mathrm{c}}> \varepsilon^*_{2\mathrm{c}}$, the brush height monotonically increases with the molfraction $x_{1\mathrm{b}}$. However, if the component 1 is a worse solvent than the component 2  ($\varepsilon^*_{1\mathrm{c}}<\varepsilon^*_{2\mathrm{c}}$) the brush height decreases with an increase of the concentration of the first component. Of course, under the assumed conditions, for  $\varepsilon^*_{1\mathrm{c}}=\varepsilon^*_{2\mathrm{c}}$ the brush height is independent of the fluid composition.

We have also  investigated the effect of interactions in the liquid mixture on the brush structure.  We have assumed that $\varepsilon^*_{1\mathrm{s}}=\varepsilon^*_{k\mathrm{s}}=\varepsilon^*_{k\mathrm{c}}
=\varepsilon^*_{22}=1$, $(k, l=1,2)$ and we have varied the parameters $\varepsilon^*_{11}$ and $\varepsilon^*_{12}$.

\begin{figure}[!t]
\centerline{\includegraphics[width=0.55\textwidth]{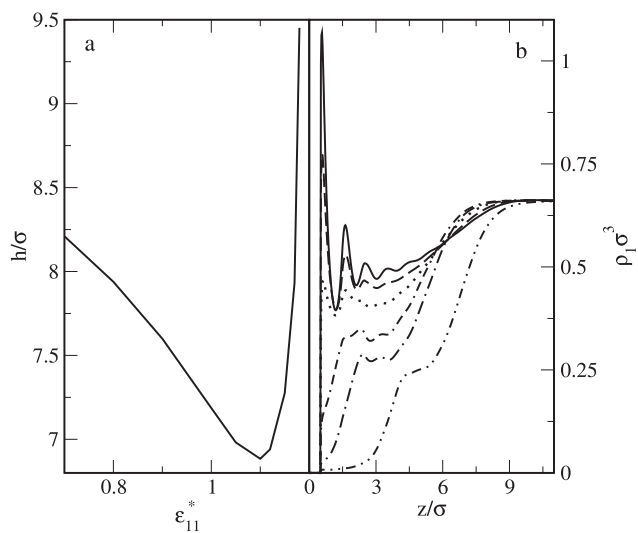}}
\caption{Part (a) shows the dependence of the brush height on the parameter $\varepsilon^*_{11}$ characterizing 11-interactions. Part (b) presents the density profiles of the fluid for selected values of the parameter $\varepsilon^*_{11}$: 0.8 (solid line), 0.9 (dashed line), 1.0 (dotted line), 1.1 (dot-double dashed line), 1.15 (dot-dashed line), 1.18 (double dotted-dashed line). The calculations were carried out for the pure component 1 and $M=18$, $\varepsilon^*_{1\mathrm{s}}=1.0$ and $\rho_\mathrm{c}^*=0.1$.} \label{fig5}
\end{figure}

\begin{figure}[!b]
\centerline{\includegraphics[width=0.48\textwidth]{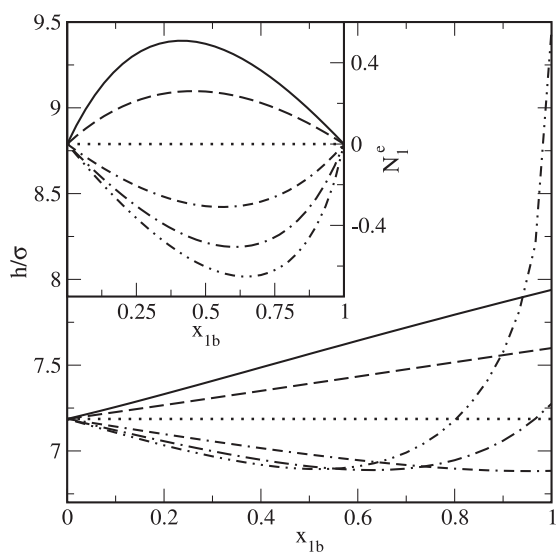}}
\caption{The dependence of the brush height on the molfraction of the first mixture component for different values of the parameter $\varepsilon^*_{11}$ characterizing interactions of the molecules: 0.8 (solid line), 0.9 (dashed line), 1.0 (dotted line), 1.1 (dot-double dashed line), 1.15 (dot-dashed line), 1.18 (double dotted-dashed line). In the inset the corresponding relative excess isotherms of the first component are shown. The calculations were carried out for $\varepsilon^*_{12}= \sqrt{\varepsilon^*_{11}\varepsilon^*_{22}}$, $M=18$, $\varepsilon^*_{1\mathrm{s}}=1.0$  $\rho_\mathrm{c}^*=0.1$.} \label{fig6}
\end{figure}

We start with the analysis of the structure of the brush immersed by a pure component 1 (see figure~\ref{fig5}). Initially, i.e., for low values of $\varepsilon^*_{11}$, the brush height decreases with an increase of the parameter
$\varepsilon_{11}^*$ to a certain minimal value. Then, the brush height rapidly increases (figure~\ref{fig5}~(a)).  These effects can be explained as follows. For  $ \varepsilon^*_{11}<1$ the interactions in the bulk fluid are weaker than the interactions of fluid molecules with the brush segments and with the underlying substrate. Therefore, the fluid molecules penetrate the brush and inhibit the coiling of the chains. Thus, the lower is $\varepsilon^*_{11}$ the higher is the  brush. When $\varepsilon^*_{11} >1$, the interactions in the bulk fluid are stronger than those with a modified adsorbent, and the fluid molecules are sucked from the surface layer. In figure~\ref{fig1}~(b) we display the density profiles of the fluid 1 for different values of the energy parameters $\varepsilon^*_{11}$.  When $\varepsilon^*_{11}<1$, the fluid molecules penetrate the brush to a great extent. The corresponding density profiles have two well-pronounced peaks near the surface.  A decrease of the parameter $\varepsilon^*_{11}$ causes an accumulation of fluid molecules in the vicinity of the surface. That is why the chains are pushed from the surface and the brush height increases. A situation changes for $\varepsilon^*_{11}>1$. In this case, depletion in the fluid density is observed close to the wall. The ``pushing effect'' gradually diminishes and the minimum height of the brush is attained for $\varepsilon^*_{11}\approx1.10$.  A different phenomenon is observed for higher values of the parameter $\varepsilon^*_{11}$. Namely, the fluid molecules located on top of the brush attract the chains and pull them in the direction of the bulk phase. This leads to the brush expansion.

\begin{wrapfigure}{i}{0.5\textwidth}
\centerline{\includegraphics[width=0.47\textwidth]{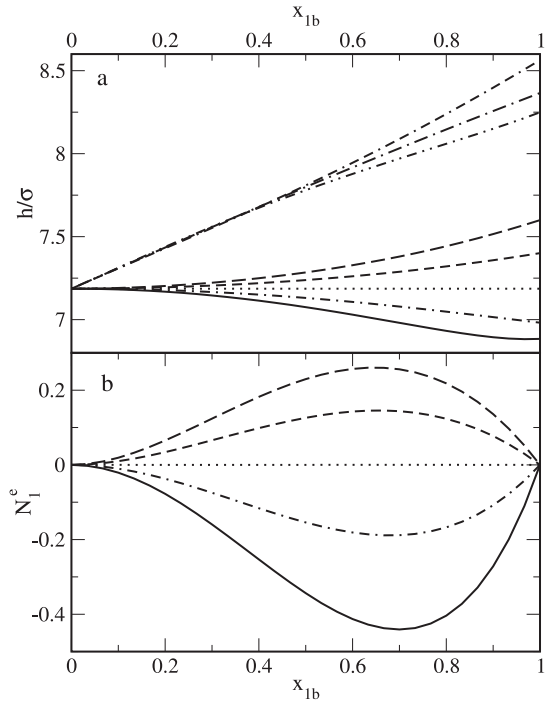}}
\caption{Part (a) -- The dependence of the brush height on the molfraction
of the first mixture component for different values of the parameters
$\varepsilon^*_{1c}$ and  $\varepsilon^*_{11}$.  The presented results
correspond to: $\varepsilon^*_{1c}=1.0$ and $\varepsilon^*_{11}=0.9$
(dashed line), 0.95 (short dashed line), 1.0 (dotted line), 1.05
(dot-short dashed line), 1.1 (solid line);  $\varepsilon^*_{1c}=1.1$ and
$\varepsilon^*_{11}=0.9$ (dot-double dashed line), 1.0 (dot-dashed line),
1.1 (double dotted-dashed line). Part (b) -- The relative excess adsorption
isotherms of the first component for $\varepsilon^*_{1c}=1.0$. The
nomenclature of the lines is the same as in the part (a). All calculation
were carried out for  $M=18$, $\varepsilon^*_{1s}=1.0$,
$\varepsilon^*_{12}=1.0$ and $\rho_c^*=0.1$.} \label{fig7}
\end{wrapfigure}

\begin{wrapfigure}{i}{0.5\textwidth}
\centerline{\includegraphics[width=0.47\textwidth]{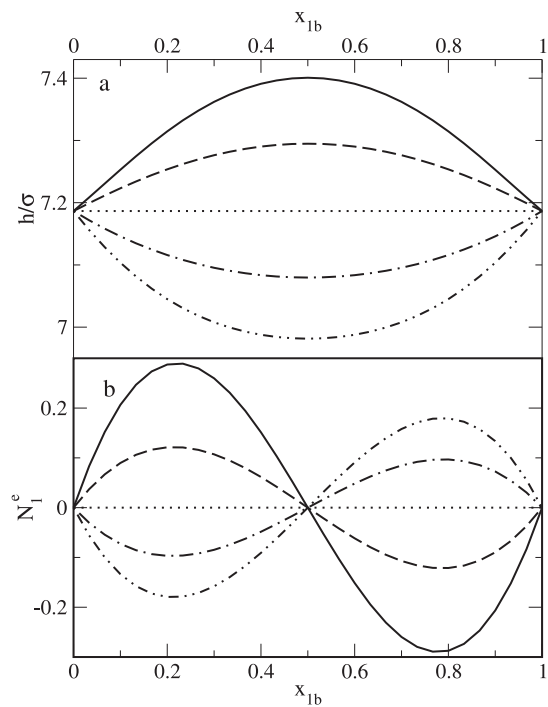}}
\caption{Part (a) --- The dependence of the brush height on the molfraction of the first mixture component for different values of the parameter  $\varepsilon^*_{12}$: 0.9 (solid line), 0.95 (dashed line), 1.0 (dotted line), 1.05 (dot-dashed line), 1.1 (double dotted-dashed line). Part (b) --- The relative excess adsorption isotherms of the first component. The nomenclature of the lines is the same as in the part~(a). All calculations were carried out for  $M=18$, $\varepsilon^*_{1\mathrm{s}}=1.0$, $\varepsilon^*_{11}=1.0$ and $\rho_\mathrm{c}^*=0.1$.} \label{fig8}
\end{wrapfigure}

The results obtained for the model mixtures are shown in figures~\ref{fig6}--\ref{fig8}.   We begin with the discussion of the behavior of the mixtures for which the Berthelot rule is fulfilled, i.e., when $\varepsilon^*_{12}=\sqrt{\varepsilon^*_{11}\varepsilon^*_{22}}$ (see figure~\ref{fig6}). In this case, at a given molfraction $x_{1\mathrm{b}}$, the parameter $\varepsilon^*_{1\mathrm{b}}$, characterizing interactions between molecules of the component, is only a factor that controls the system behavior. For  $\varepsilon^*_{11}<1$, the brush height is an increasing function of the molfraction of the first component. In such systems the interactions of the fluid molecules with the solid surface and with the grafted chains are stronger than the interactions between molecules in the bulk mixture. The tethered chain is much more extended in the first component than in the second. The brush height is a monotonically increasing function of the molfraction $x_{1\mathrm{b}}$. If   $\varepsilon^*_{11}=1$, the brush height does not depend upon the mixture composition.  The effect of the parameter $\varepsilon_{11}^*$ on the brush height is more complicated for   $\varepsilon^*_{11}>1$. When the 11-interactions are slightly stronger than the interactions with the bare surface and with segments of grafted chains, the brush height decreases with an increase of the molfraction $x_{1\mathrm{b}}$. However, for $\varepsilon^*_{11}=1.10$ and $\varepsilon^*_{11}=1.18$ there is a minimum on the curve $h~\text{vs}~x_{1\mathrm{b}}$. At high concentrations of the first component, the brush height  rapidly increases. Such systems behave similarly to the grafted chains in contact with the pure component 1 described above.

The interactions in the mixture can be also characterized by the exchange parameter $\omega_{12}^*= \varepsilon_{12}^*- 0.5\left(\varepsilon_{11}^*+\varepsilon_{22}^*\right)$. In the studied systems $\omega_{12}^*<0$ for $\varepsilon_{11}^*<1$ and $\omega^*_{12}>0$ for $\varepsilon_{11}^*>1$.
In the inset we show the relative excess isotherms corresponding to the curves  $h$ vs $x_{1\mathrm{b}}$. We see that the relative excess adsorption of the first component is positive for $\omega_{12}^*<0$ and negative for $\omega_{12}^*>0$.

We have also considered the model mixtures for which the parameters $\varepsilon^*_{11}$ and $\varepsilon^*_{12}$ can be changed independently. This allows us to separate the effects connected with  interactions between the same molecules ($\varepsilon^*_{11}$) and between unlike molecules ($\varepsilon^*_{12}$) in the mixture. First, we have calculated the brush height for  $\varepsilon^*_{12}=1$,  two values of the parameter $\varepsilon^*_{1\mathrm{c}}=1$ and $\varepsilon^*_{1\mathrm{c}}=1.1$ and different values of the parameter $\varepsilon^*_{11}$.  In the both cases, if  $\varepsilon^*_{11}<1$, the brush height increases with an increase of the concentration of the first component in the mixture and the opposite effect is observed for $\varepsilon^*_{11}>1$ (figure~\ref{fig7}~(a)). Notice that the affinity of the first component to the grafted chains $\varepsilon_{\mathrm{c}1}^*$ affects the brush height more strongly than the parameter  $\varepsilon_{11}^*$. The relative excess isotherms corresponding to the curves $h~\text{vs}~x_{1\mathrm{b}}$ calculated for $\varepsilon_{\mathrm{c}1}^*=1$ are shown in part (b). These results are analogous to those presented in figure~\ref{fig6}.

Now, we discuss the results obtained for symmetrical mixtures. We have assumed that $\varepsilon^*_{11}=\varepsilon^*_{22}$ and varied the parameter $\varepsilon^*_{12}$. The selected results are shown in figure~\ref{fig8}. In this case the plots $h~\text{vs}~x_{1\mathrm{b}}$ have regular shapes, namely, there is one extreme at  $x_{1\mathrm{b}}=0.5$ on each curve. This is a maximum for $\varepsilon_{12}^*<1$ ($\omega_{12}^*<0$) and a minimum for $\varepsilon^*_{12}>1$  ($\omega_{12}^*>0$) (see figure~\ref{fig8}~(a)). The relative excess isotherm $N^\mathrm{e}_1$ has an azeotropic point, i.e., the point at which the relative adsorption isotherm is equal to zero and  $0<x_{\mathrm{az}}<1$, just at  $x_{1\mathrm{b}}=0.5$. The relative excess adsorption isotherm changes the sign at the azeotropic point. When  $x_{1\mathrm{b}}< x_\mathrm{az}$, the relative excess isotherm $N^\mathrm{e}_1$ is positive for $\omega_{12}^*<0$ and negative for $\omega_{12}^*>0$.  In the case of stronger deviation of the exchange parameter from $\omega_{12}^*=0$ the shapes of the curves $h~\text{vs}~x_{1\mathrm{b}}$ can be different. This problem will be a subject of investigations in the future.

In summary, the structure of the surface layer varies considerably with the change of the composition of the mixture. The grafted chains can be more or less extended in the mixtures of different compositions. The curves describing the brush height versus the molfraction of a given component have various shapes depending on the relation between the parameters characterizing the system. When the components have different affinities to the underlying surface and to the grafted chains, the mixture composition  strongly affects the brush height. We conclude that stronger interactions of the fluid molecules with the wall cause an increase of the brush height. The same trend is observed in the case of the interactions with the segments of grafted chains. The nature of liquid mixture also plays a significant role in the process of coiling of the grafted chains. Interactions between molecules of particular components of the mixture effect their adsorption on the bonded-phase, which, in turn, affects the height of the brush. The different dependences of the brush height on the mixture composition are observed for different sets of parameters of the model. The grafting density is an important parameter that can decide about the brush height. We have analyzed the dependence of the brush height on the grafting density for different mixtures. We have found that the scaling relation describing the changes in the brush height with the changes of the grafting density is approximately satisfied within the region of stretched chains and the exponent $\gamma$ depends on the type of mixture and its compositions. Our findings are qualitatively consistent with the computer simulation carried out for selected systems~\cite{c28,c29,c37}. Unfortunately, there are no systematic simulation studies of the effect of mixture composition on the structure of the layer of the grafted chains. The problem of comparing the theoretical predictions with the simulation data requires further investigations.

\section*{Acknowledgements}

This work is supported by the Ministry of Science of Poland under the Grant No. N N204151237 and by EC under the Grant No.~PIRES--GA2010--268498 (MB).

\ukrainianpart

\title%
{Дослідження методом функціоналу густини структури прив'язаних ланцюжків у бінарній суміші%
}
\author[]{М. Борувко, T. Сташевскі}
\address{Відділ моделювання фізико-хімічних процесів, університет Марії Кюрі-Склодовської, \\ 20031 Люблін, Польща}

\makeukrtitle

\begin{abstract}
\tolerance=3000%
Метод функціоналу густини використовується з метою дослідження
структури шару, утвореного ланцюжковими молекулами, які прикріплені до
твердої поверхні. Ланцюжки моделюються як вільно зв'язані
сфери. Сегменти ланцюжків і всі компоненти взаємодіють за допомогою
потенціалу Леннарда-Джонса (12--6).  Взаємодії молекул плину зі стінкою
описуються потенціалом Леннарда-Джонса (9--3). Ми аналізуємо як різні
параметри моделі впливають на залежність висоти щітки від концентрації
суміші. Ми розглядаємо вплив густини прищеплення і параметрів, що
характеризують взаємодії молекул плину з підкладкою і з ланцюжками, як
і з взаємодіями в суміші. Зміни висоти щіток корелюють з адсорбцією
окремих компонент.
\keywords щітки, адсорбція, теорія функціоналу густини
\end{abstract}

\end{document}